\RequirePackage{fix-cm}
\documentclass [smallextended, twocolumn] {svjour3} [natbib]

\smartqed  

\usepackage{graphicx,latexsym}
\usepackage{amssymb,amsmath,bm}
\usepackage{subfigure}
\usepackage{braket}
\usepackage{multicol}
\usepackage{siunitx}
\usepackage{array}
\usepackage{float}
\usepackage[nopar]{lipsum}
\usepackage{caption}
\usepackage{multirow}
\usepackage[numbers, sort & compress]{natbib}
\usepackage{bigstrut}
\usepackage[super]{nth}

\newcommand{\angstrom}{\text{\normalfont\AA}}
\usepackage{hyperref}
\hypersetup{
	pdfnewwindow=true,       
	colorlinks=true,         
	linkcolor=blue,          
	citecolor=black,          
	filecolor=magenta,       
	urlcolor=black           
}

\usepackage[normalem]{ulem}

\def\sec#1{Sec.\ \ref{#1}}

\def\fig#1{Fig.\ \ref{#1}}

\begin{document}

\title{Properties of bilayer graphene-like Si$_{2}$C$_{14}$ semiconductor using first-principle calculations
}
\subtitle{ }


\author{Nzar Rauf Abdullah      
}


\institute{Nzar Rauf Abdullah  \at
              {Division of Computational Nanoscience, Physics Department, College of Science, 
              	University of Sulaimani, Sulaimani 46001, Kurdistan Region, Iraq\\ 
              	Computer Engineering Department, College of Engineering, Komar University of Science and Technology, Sulaimani 46001, Kurdistan Region, Iraq\\
              	Science Institute, University of Iceland, Dunhaga 3, IS-107 Reykjavik, Iceland 
              Tel.: +770 144 3854\\
              \email{nzar.r.abdullah@gmail.com}           
 }
}

\maketitle

\begin{abstract}
We theoretically investigate silicon doped bilayer graphene, Si-BLG, with different stoichiometry of Si-BLG structures. The dangling bonds of C-Si atoms are found at low concentration ratio of Si atoms inducing sp$^3$-hybridization of buckled pattern in the structure. 
The Si-BLG with dangling bonds revel strong mechanical response at low strain of a uniaxial load, and 
the fracture strain is seen at low strain ratio. The sp$^3$-hybridization forms a small bandgap which induces an intermediate thermal and optical response of the system. 
In contrast, at higher Si concentration ratio, the Young modulus and fracture strain are increased comparing to the low Si concentration ratio. This is due to presence of high number of C-Si bonds which have a high tolerant under uniaxial load.
In addition, a relatively larger bandgap or an overlap between valence and conduction bands are found depending on the Si configurations. In the presence of gaped Si-BLG, the thermal and optical response are high. We thus obtain a high Seebeck coefficient and a figure of merit with low electronic thermal conductivity which are useful for thermoelectric nanodevices, and an enhancement of optical response is acquired with a redshift in the visible range.
\end{abstract}

\section{Introduction} 

The ability to fabricate of monolayer graphene \cite{Novoselov666} leads researchers to 
study quantum Hall effect plateaus that is explained in terms of Dirac-like chiral quasiparticles 
with Berry phase $\pi$. Consequently, bilayer graphene (BLG) has been considered as one of the interested structure of study \cite{MCCANN2007110, Nzar.25.465302}.
One of the interested feature of BLG is the possibility to locally induce a bandgap and tune its magnitude by applying a strong electric field perpendicular to the carbon nanosheets \cite{Kanayama2015, 10.3389/fphy.2014.00003, PhysRevB.95.195307, ABDULLAH2018223}. This property of BLG can be used to design next-generation transistors that would work faster and use less energy \cite{7292288, 7736978}.

The physical properties such as the bandgap of BLG can also be tuned by doping of foreign atoms 
\cite{doi:10.1021/nn202463g, ABDULLAH2020126807}. Several foreign atoms have been used to improve physical properties of monolayer and bilayer graphene such as Boron/Nitrogen (B/N) atoms \cite{NEMNES2018175, HAN2015618,ABDULLAH2020126350, ABDULLAH2020126807}, cesium (Cs) \cite{doi:10.1021/acsnano.9b08622, abdullah2019thermoelectric}, Silicon (Si) atom \cite{DENIS2010251, ABDULLAH2016280},  and transition metals \cite{C6CP01841F}. The BLG with B/N doping exhibits a semiconding material with a small bandgap, 
where the Fermi energy is located in the bandgap \cite{doi:10.1002/adma.200901285, ABDULLAH2020103282} which can be used 
to study photocatalysis \cite{doi:10.1063/1.4950993}. The tunning bandgap by Si doped BLG have been reported and shown that the Si atoms induces a small bandgap due to the silimar valence electron of Si and C atoms \cite{DENIS2010251, gudmundsson2019coexisting}.
The sandwiching a graphene by Cs induces 
thermodynamically stable flat band materials \cite{Hase2018}. In the the transition metals doped BLG, the electronic and magnetic properties as a function of strain have been also analyzed to indicate that the strain is important for the stabilities of the high-coverage TM-intercalated BLG \cite{C6CP01841F,TANG20171529, en12061082, ABDULLAH20181432}. The bandgap tuning of BLG can thus improve the electrical, the thermal and the optical conductivities and modulate the mechanical properties \cite{McCann_2013}. 

The Si doping monolayer graphene has been investigated by several research groups. 
The Si atoms influences mechanical property such as Young’s modulus of monolayer graphene, the 
optical \cite{Houmad2015} and  thermal \cite{C5NR06345K, RASHID2019102625, Abdullah2019, Fatah_2016} characteristics. 
But the electronic, mechanical, thermal, and optical properties of Si-doped BLG have not 
been systematically investigated. 
In this work, we consider Si-doped AA-stacked BLG represented by SiC$_{15}$, and Si$_{2}$C$_{14}$ structures depending on Si concentrations. The electronic, mechanical, optical and thermal characteristics are investigated using density functional theory. A comparison for different Si atoms configurartion in BLG will be shown with detailed mechanisms of improving the BLG structures by Si impurity atoms \cite{abdullah2019manifestation, doi:10.1063/1.4904907, TANG20173960, ABDULLAH2020114221}.

In \sec{Sec:Model} the structure of BLG is briefly over-viewed. In \sec{Sec:Results} the main achieved results are analyzed. In \sec{Sec:Conclusion} the conclusion of the results is presented.

\section{Computational details}~\label{Sec:Model}
The calculations performed in this study are using DFT within Generalized Gradient Approximation (GGA) \cite{PhysRevLett.77.3865} and the computer software is Quantum Espresso (QE) package \cite{Giannozzi_2009}. 
The calculations are also carried out on a $2\times2$ bilayer supercell arrangement with plane-wave basis set of Perdew-Burke-Ernzerhof (PBE) psuedopotentials \cite{PhysRevB.23.5048} with $1088.45$~eV cutoff-energy. Under the GGA-PBE, the van der Waals interactions between the layers of system is almost ignored.
The structure relaxation is performed using the $k$-point grid is $12\times12\times1$ and 
the force on each atom is less than $10^{-6}$ eV/$\angstrom$.
For the Brillouin zone sampling and the calculations of the density of state (DOS), a $12\times12\times1$ and a $77\times77\times1$ grids are used, respectively.

The XCrySDen, crystalline and molecular structure visualization program, 
is uilized to visualize our structures \cite{KOKALJ1999176, Nzar.25.465302, Nzar_ChinesePhysicsB2016}. 
Furthermore, the thermoelectric properties of the systems are investigated using the Boltzmann transport software package (BoltzTraP)~\cite{madsen2006boltztrap-2}. 
The BoltzTraP code uses a mesh of band energies and has an interface to the QE package \cite{ABDULLAH2020126578, ABDULLAH2020113996}. The optical peroprties of the systems are obtained 
by the QE code with the broadening of $0.5$~eV.

\section{Results}~\label{Sec:Results}

In this section, we show the results of structural stability, electronic, mechanical, thermal, and optical properties of the Si-BLG with different concentration and configurations of Si atoms in a 
$2\times2$ bilayer supercell \cite{ABDULLAH2019102686}.

\subsection{Model}

\begin{figure}[htb]
	\centering
	\includegraphics[width=0.45\textwidth]{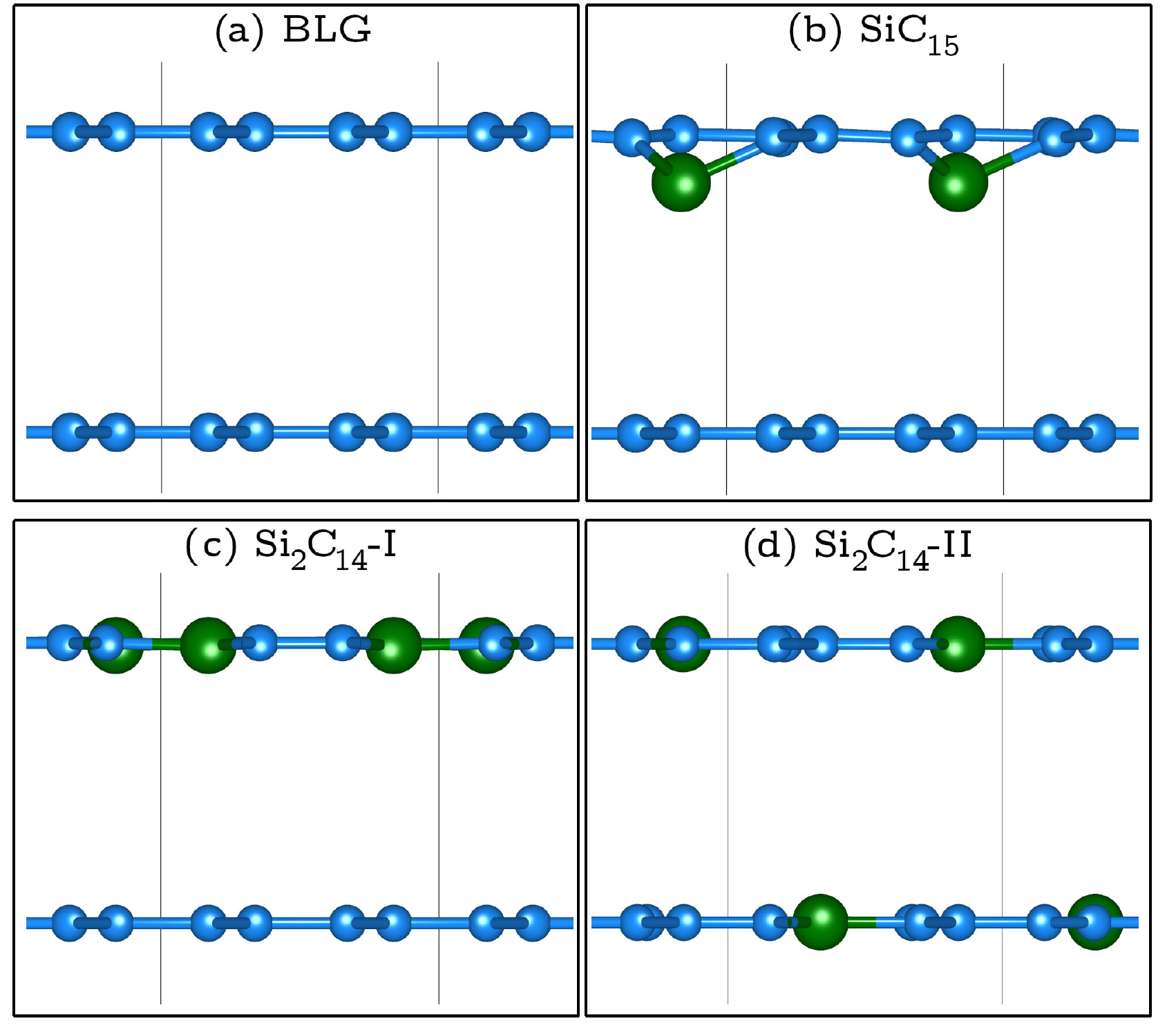}
	\caption{AA-stacked pristine BLG (a), SiC$_{15}$ (b), Si$_2$C$_{14}$-I (c), and Si$_2$C$_{14}$-II (d). The C and Si atoms are blue and green colors, respectively.}
	\label{fig01}
\end{figure}

In \fig{fig01} the prisine BLG (a) and Si-BLG (b-d) structure are presented.  
We consider one Si atom is doped in the top layer with the doping 
concentration ratio of $6.25\%$ identifying as SiC$_{15}$ (b). The vertical gray lines indicate 
the border of $2\times2$ supercell \cite{JONSSON201781}.
In addition, we also assume two Si atoms doped in BLG with doping concentration ratio of $12.5\%$ in two configurations of Si atoms: First, both Si atoms are doped in the top layer at the para-positions identifying as Si$_2$C$_{14}$-I (c). Second, one Si atoms is doped at the para-position of the top layer and another Si atom is put at meta-position of the bottom layer called Si$_2$C$_{14}$-II (d). 
We avoid the formation of Si-Si bonds in the Si-BLG, which substantially destabilize the hexagonal system \cite{Li2014}.

The interlayer distance, lattice constant, and the C-C bond length of pristine BLG are found to be  $4.64$, $2.46$, $1.42$~$\angstrom$, respectively, which are expected for the 
GGA-PBE calculations. But if the van der Waals interactions in the exchange (XC)
functional with LDA is included, the interlayer distance becomes $3.6$~$\angstrom$, which is close to experimental work \cite{ABDULLAH2020100740, doi:10.1063/1.2975333, PhysRevB.82.195325, nzar_2019_Annalen}. 
The interlayer distance for all three Si-BLG structures is $4.64$~$\angstrom$ which is almost unchanged indicating no interlayer interaction energy due to Si atoms. 
This is opposite to the interalayer repulsive interaction between the Si-Si atoms in BLG which is recently observed when the LDA exchange correlation is assumed \cite{abdullah2020interlayer, doi:10.1021/acsphotonics.5b00115}. 

Another point is that the Si atom in top layer of SiC$_{15}$ is moved towards the bottom layer with 
$0.7$~$\angstrom$ away from the top layer. This deviation of Si atoms from graphene surface has been 
reported with same deviation length, $0.7$~$\angstrom$~\cite{DENIS2010251,abdullah2020properties}. 
The Si shifting here forms the dangling bonds of Si-C in which the Si atom tends to adopt sp$^3$ hybridization, shuch as in silicene where the buckled patter forms. 
It thus increases the C-Si bond length to $1.71$~$\angstrom$ and slightly increase the average C-C bond lengths to $1.45$~$\angstrom$.
This shifting of the Si atom will influence the physical properties of the system as it will be shown layer.

In the Si$_2$C$_{14}$-I, the average C-C bond length in the top(bottom) layer is $1.37$($1.53$)~$\angstrom$, and the C-Si bond length is $1.69$~$\angstrom$. 
In the case of Si$_2$C$_{14}$-II, the average C-C bond length in the top(bottom) layer is $1.47$($1.5$)~$\angstrom$, and the C-Si bond length is $1.68$~$\angstrom$. It can be clearly seen that the position of Si atoms in the hexagonal structure of graphene influence the bond length and thus the physical properties of the systems \cite{https://doi.org/10.1002/andp.201600177, Rauf_Abdullah_2016, https://doi.org/10.1002/andp.201700334}.

\subsection{Mechanical response}

\begin{figure}[htb]
	\centering
	\includegraphics[width=0.35\textwidth]{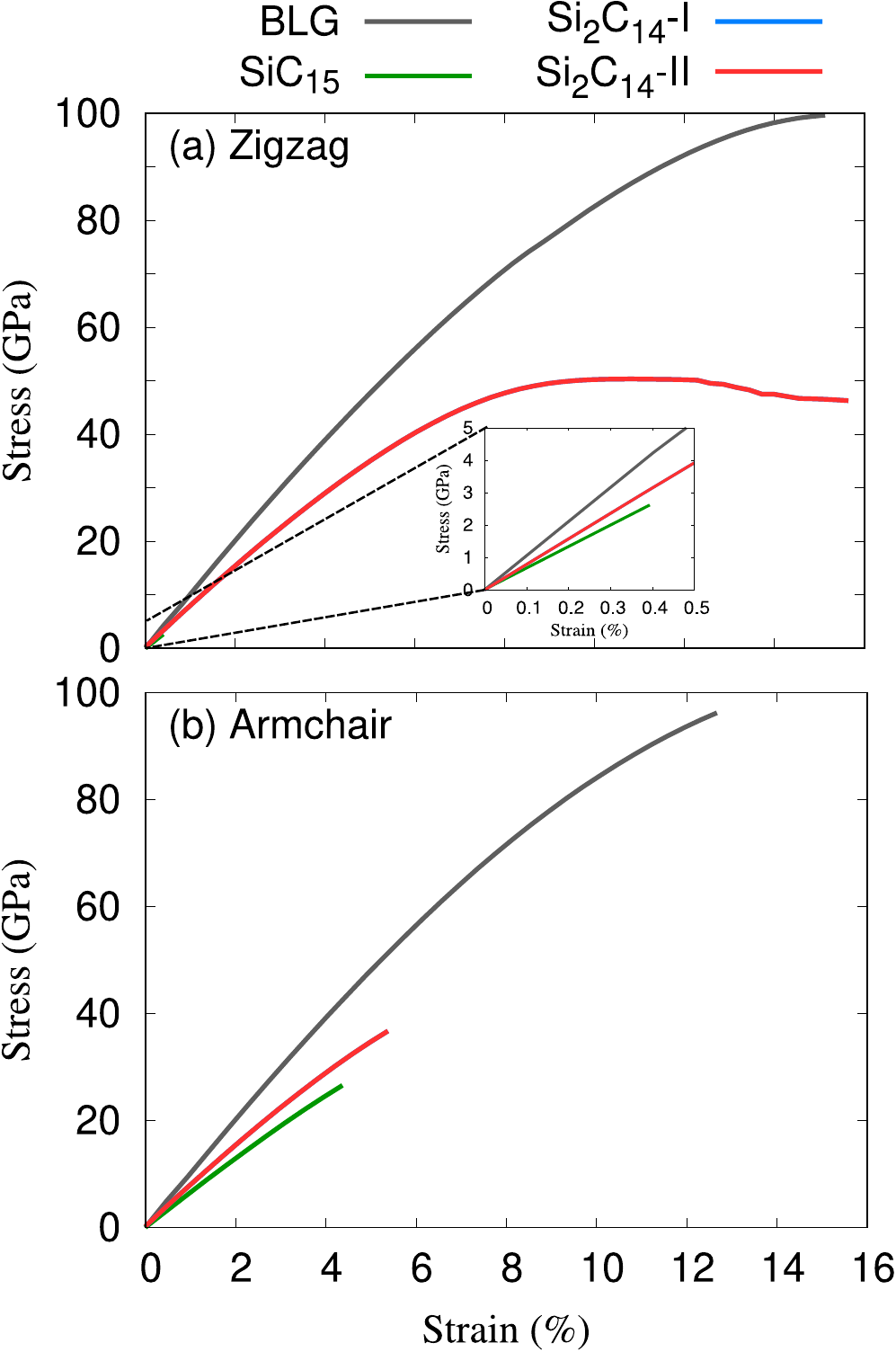}
	\caption{Stress-strain curves for pristine BLG (gray), SiC$_{15}$ (green), Si$_2$C$_{14}$-I (blue), and Si$_2$C$_{14}$-II (red) in the zigzag (a) and armchair (b) directions.}
	\label{fig02}
\end{figure}

The modifications of bond lengths and the Si-BLG structures affect the mechanical properties of the system. The stress-strain curves for pristine BLG (gray), SiC$_{15}$ (green), Si$_2$C$_{14}$-I (blue), and Si$_2$C$_{14}$-II (red) in the zigzag (a) and armchair (b) directions are shown in \fig{fig02}.
In 2D materials, the two basis vectors are perpendicular to each other. We therefore apply 
uniaxial strains directly along the vector direction. 
Tension simulation is carried out by imposing in-plane tensile strain at a rate of 0.02
per step in the uniaxial zigzag or armchair direction. 

The stress-strain curves of pristine BLG are consistent with those existing in the literature for both zigzag and armchair directions, showing the reability of our calculations \cite{doi:10.1063/1.4789594}. Similar to previous investigations, the pristine BLG obeys linear relationships within a small strain range up to $5\%$. The linear elastic region gives the Young modulus of $974$~GPa which is very close to experimental value \cite{Lee385}. In addition, the ultimate strength of pristine BLG is $99.64$~GPa (zigzag) and $96.22$~GPa (armchair) at atrain of $0.151$ and $0.126$, respectively. Since there is no stretching in the pristine BLG after the fracture
strain, the values of ultimate stress are also equal to the fracture strain.
This slightly anisotropy of the ultimate stress in zigzag and armchair directions has also been observed for pristine monolayer graphene~\cite{Lu_2018}.

In the Si-BLG structures, it is interesting to see that linear elastic region and ultimate strength for SiC$_{15}$ are decreased in both zigzag and armchair directions. The reduction is much stronger in the zigzag direction (see \fig{fig02}(inset)). The ultimate strength is $3.8$~GPa with strain $0.4\%$ in the zigzag direction, and $26.58$~GPa at strain $4.3\%$ in the armchair direction.
This is attributed to the dangling bonds of Si-C in which the Si atom tends to adopt sp$^3$ hybridization. In addition, the moving out of Si atom introduces extremely unbalanced bonds strength which distorts the perfect hexagon rings of the $2\times2$ supercell honeycomb. 
Consequently, the pre-elongated C-C and Si-C bonds of SiC$_{15}$ under the uniaxial tension are lifted. 

One also can clearly see that the stress-strain curves for both Si$_2$C$_{14}$-I, and Si$_2$C$_{14}$-II in the zigzag and armchair directions are irrespective of doping configurations, and
they are the same because number of Si-C bonds are equal with different configurations in both structures. The Yong modulus and tensile or ultimate strength are found to be $732$($728$), and $50.22$($36.72$) for the zigzag(armchair) directions, respectively.
The reduction of stress-strain curves for Si$_2$C$_{14}$-I, and Si$_2$C$_{14}$-II refers to the presence of more Si-C bonds in these structures.  
In addition, the ultimate strength and fracture strain in the zigzag direction is higher than that of the armchair which is due to existing of more Si-C bonds in the zigzag direction. This is because in the Si-C bonding the charge distribution of the Si atom is more easily reshaped than that of the C atom as the Si atom has much lower electronegativity. It arises Si-C bonds much more tolerance of the stretched process.

\subsection{Band energy and DOS}

\begin{figure}[htb]
	\centering
	\includegraphics[width=0.45\textwidth]{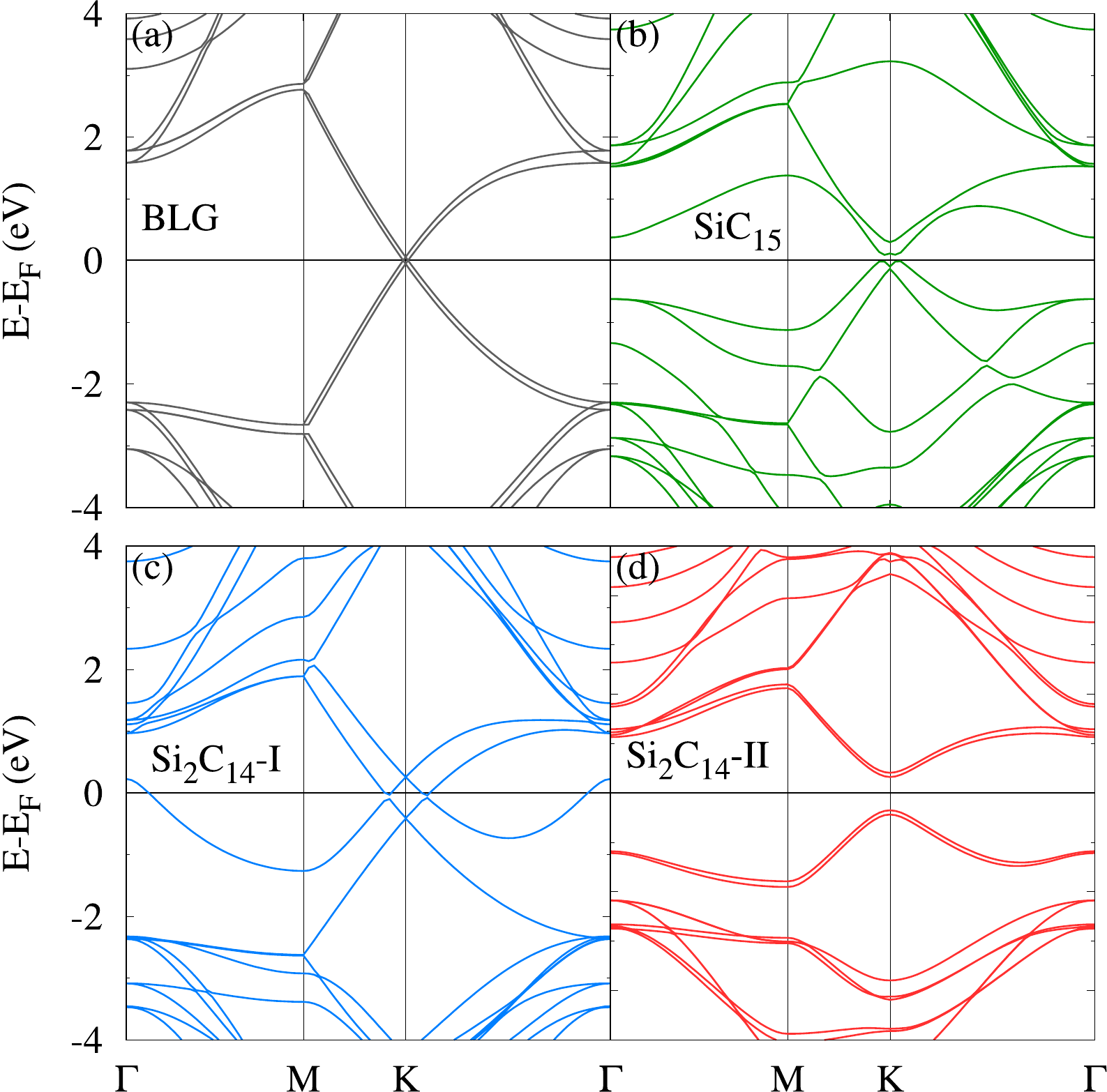}
	\caption{Electronic band structure of pristine BLG (a), SiC$_{15}$ (b), Si$_2$C$_{14}$-I (c), and Si$_2$C$_{14}$-II (d).}
	\label{fig03}
\end{figure}

\begin{figure}[htb]
	\centering
	\includegraphics[width=0.42\textwidth]{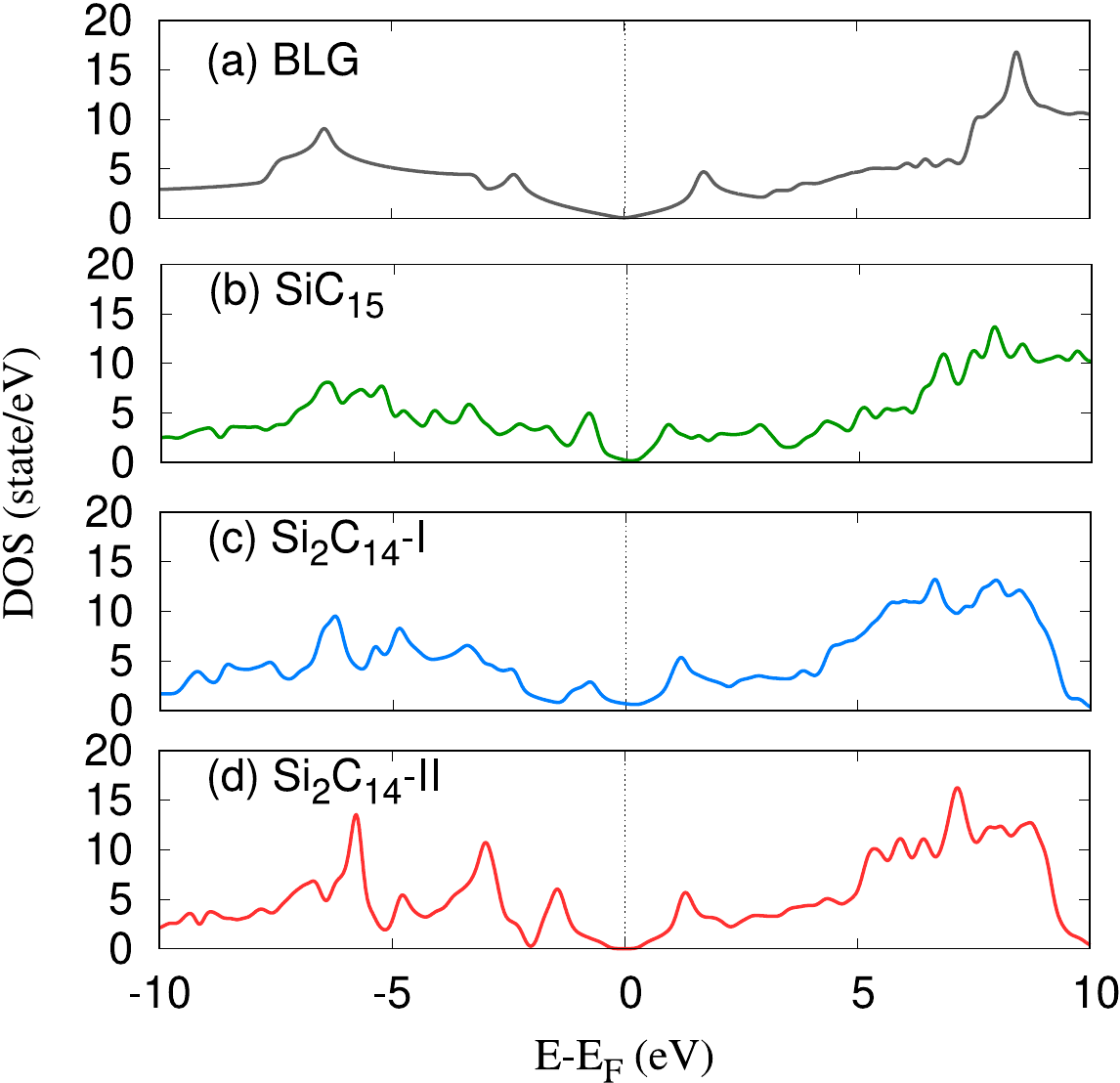}
	\caption{Density of state of pristine BLG (a), SiC$_{15}$ (b), Si$_2$C$_{14}$-I (c), and Si$_2$C$_{14}$-II (d).}
	\label{fig04}
\end{figure}

The electronic band structures and the DOS of pristine BLG (a) and Si-BLG (b-d) are shown in \fig{fig03}, and \fig{fig04}, respectively. The linear dispersion
of the first valence band, $\pi$, and the conduction band, $\pi^*$, of the pristine BLG is found with 
zero bandgao and DOS.  The energy spacing between the $\pi_2\text{-}\pi_1$($\pi^*_2\text{-}\pi^*_1$) is almost $0.11$~eV \cite{GUDMUNDSSON20181672}. 
The linear dispersion of pristine BLG band structure around the Fermi level shows the semimetallic nature where valence band maxima ($\pi$ band) and the conduction band minima ($\pi^*$ band) touch each other only at the high symmetry K point. 
The Hamiltonian that defines the electronic structure of pristine BLG around the Dirac point can be given as \cite{aliofkhazraei2016graphene}
\begin{equation}
	\hat{H} = 
	\begin{pmatrix}
		\Delta               & \hbar v_F(k_x-i k_y) \\
		\hbar v_F(k_x+i k_y) & -\Delta
	\end{pmatrix}
\end{equation}
where, $\Delta$, $v_F$, and $k$ indicate onsite energy difference between the C atoms located at A and B sites, Fermi velocity, and  momentum  of charge carriers. 
The linear relation refers to the zero value of the onsite energy difference, $\Delta$, in pristine BLG arising from the presence of inversion symmetry in pristine BLG. 
The zero value of onsite energy difference refers to the potentials
seen by the C atoms at the sites A and B are the same. Consequently, there is no opening up 
of energy gap in monolayer and bilayer graphene.

In the Si-BLG, the overall energy spacing between $\pi$ and $\pi^*$ at M-point determining the interlayer interaction is not much changed because the GGA-PBE is assumed where the interalayer interaction is ignored. But a strong influence between $\pi$ and $\pi^*$ at $\Gamma$-point is seen inflecting the optical transitions \cite{ABDULLAH2020113996,ABDULLAH2019102686, https://doi.org/10.1002/andp.201500298}. 
The bandgap of SiC$_{15}$ and Si$_2$C$_{14}$-II are found to be $0.11$ and $0.67$~eV exhibiting semiconducting materials. The small bandgap of Si doped graphene may refers to the same number of the valence electron of Si and C atoms.
In general, a band gap is opened near the K point due to breaking of
the inversion symmetry by the distortion generated by the Si atoms configuration. 
The reason for the broken symmetry is that the potential seen by the atoms at sites A and B is now different, leading to a finite value of onsite energy $\Delta \neq 0$, where  $\Delta = \alpha (V_{\rm A} - V_{\rm B})$ with $\alpha$ being a constant value and V$_{\rm A}$(V$_{\rm B}$) is the potential seen by an atom at the site A(B)~\cite{Rahman_2014}. 
Furthermore, a small overlap of valence and conduction bands of Si$_2$C$_{14}$-I indicate a metallic behavior of this structure. These modifications in electronic band structure and induced bandgap will effectively change the thermal properties of the systems.

\subsection{Thermal response}

\begin{figure}[htb]
	\centering
	\includegraphics[width=0.48\textwidth]{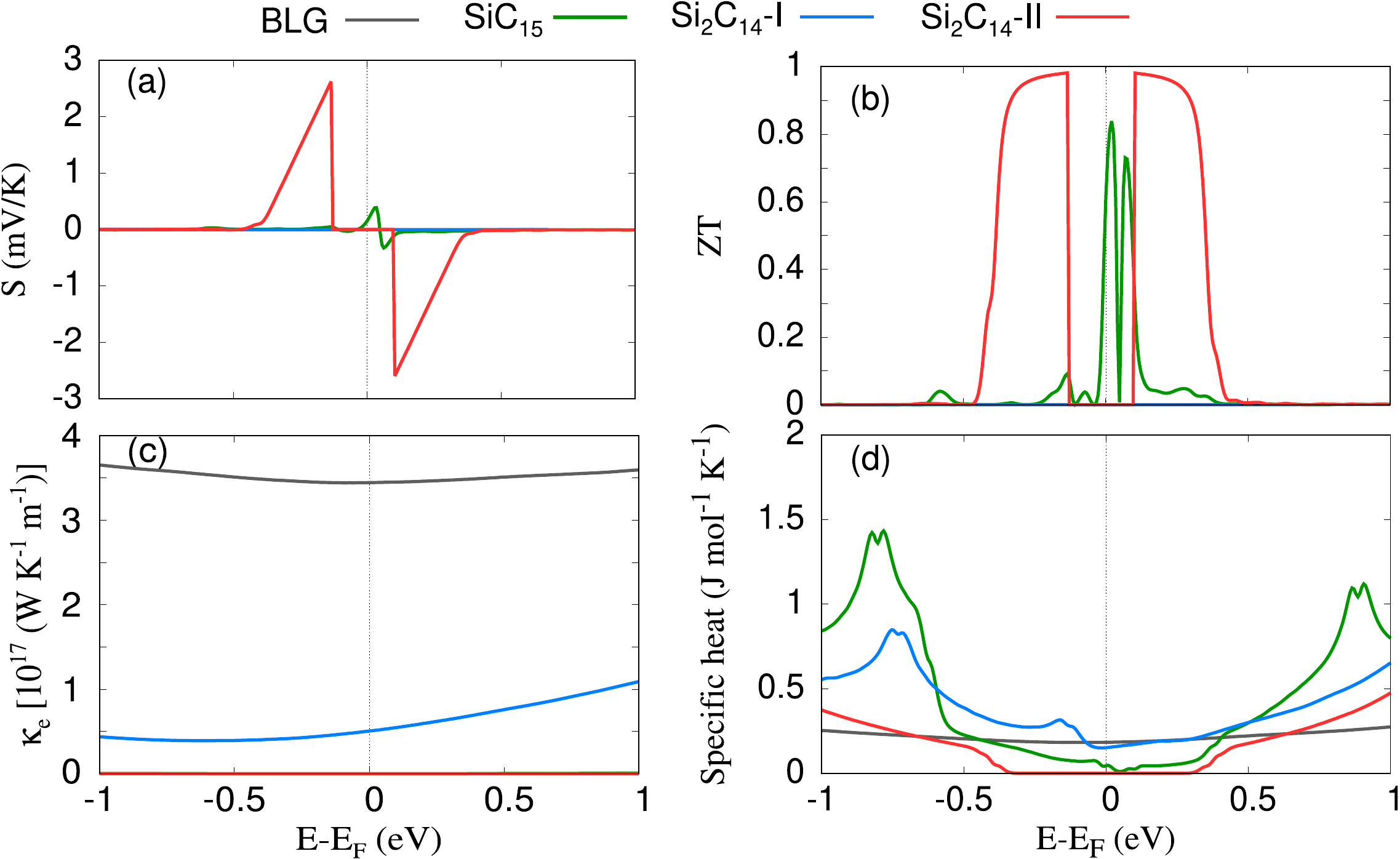}
	\caption{Seebeck coefficient, $S$, (a), Figure of merit, ZT, (b), electronic thermal conductivity, $\kappa_e$, (c), and specific heat (d) at $T = 100$~K for pristine BLG (gray), SiC$_{15}$ (green), Si$_2$C$_{14}$-I (blue), and Si$_2$C$_{14}$-II (red).}
	\label{fig05}
\end{figure}

The thermal properties such as Seebeck coefficient (a), figure of merit (b), electronic thermal 
conductivity (c), and specific heat (d) at temperature $T = 100$~K are shown in \fig{fig05} for the pristine BLG and Si-BLG structures \cite{doi:10.1021/acsphotonics.5b00532, ABDULLAH2018102, Abdullah_2018}. We are interested in thermal behavior of the systems at low temperature and it has been reported that the electrons and phonons thermal behaviors are decoupled at low temperature range from 20 to 160 K \cite{PhysRevB.87.241411}. We thus present the electronic part of thermal properties.

The $S$ and the $ZT$ of pristine BLG and Si$_2$C$_{14}$-I are very small due to the zero bandgap and overlapping the valence and conduction bands, respectively. 
This is attributed to the cancellation effect from electron-hole contributions to the transport quantities. 
So, the highest thermal conductivity and specific heat are thus found around the Fermi energy for these two structures. 
In contrast, opening up the bandgap of SiC$_{15}$ and Si$_2$C$_{14}$-II maximizes the $S$, and $ZT$ and minimizes the thermal conductivity and specific heat around the Fermi energy. Therefore, the maximum $S$ and $ZT$ are found for Si$_2$C$_{14}$-II as it has the maximum bandgap among these structures.

\subsection{Optical response}

\begin{figure}[htb]
	\centering
	\includegraphics[width=0.45\textwidth]{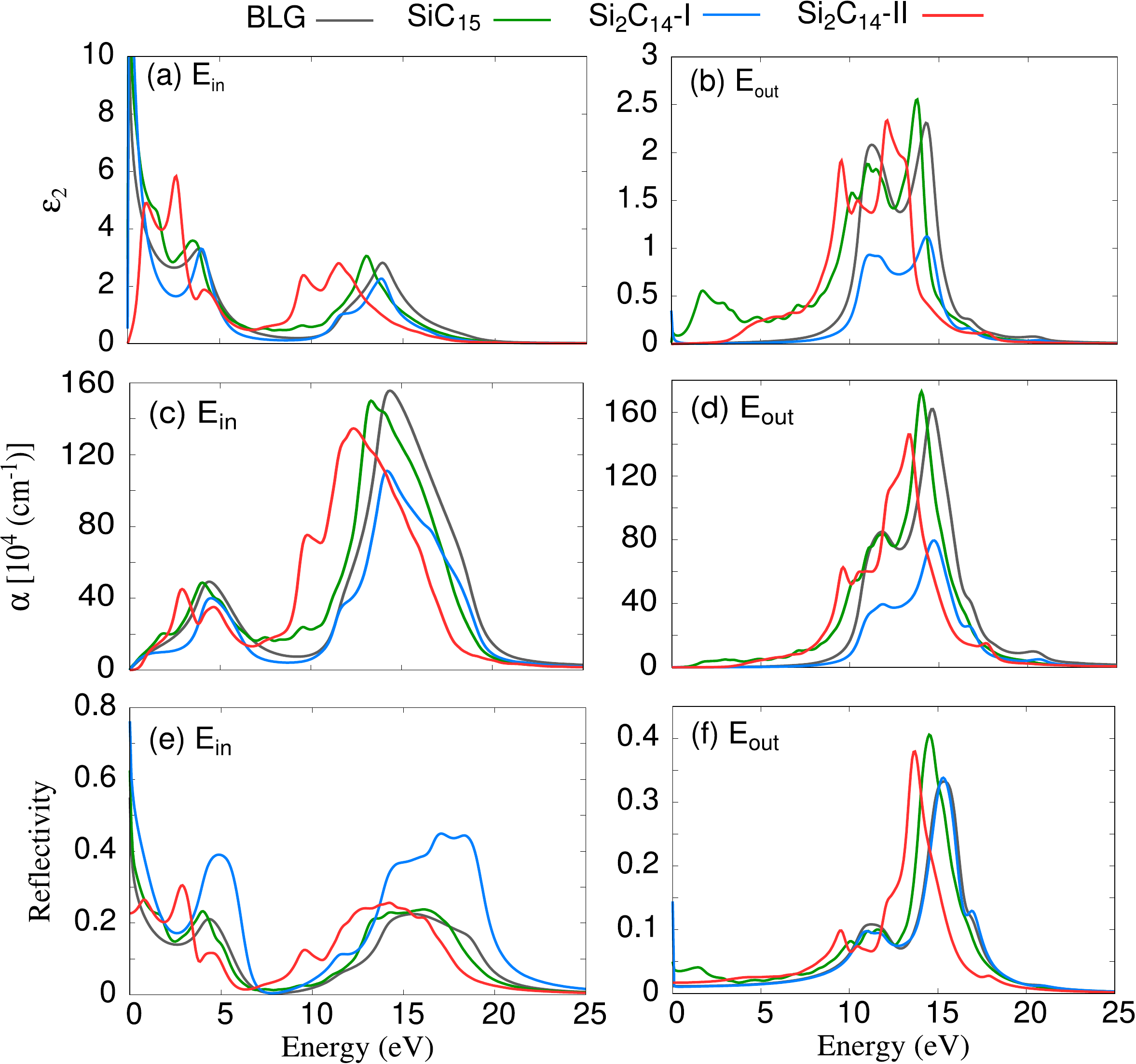}
	\caption{Imaginary part of dielectric function, $\varepsilon_2$, 
		(a-b), absorption coefficient, $\alpha$, (c-d), and Reflectivity (e-f) of pristine and Si-BLG for in-plane (left panel) and out-plane (right panel) of an applied electric fields.}
	\label{fig06}
\end{figure}

The modified electronic band structures and DOSs due to Si dopants are expected to tune the optical response of BLG.
The optical response of the Si-BLG under applied in- (left panel) and out-plane (right panel) electric fields is demonstrated in \fig{fig06}, where the imaginary part of dielectric function 
(a-b), the absorption coefficient (c-d), and the Reflectivity, $R$, (e-f) of pristine BLG and Si-BLG are presented.

In the case of pristine BLG, two main peaks in the imaginary dielectric function are observed 
at $3.95$~eV corresponding to the $\pi$ to $\pi^*$ transition and at $13.87$~eV generated due to the $\sigma$ to $\sigma^*$ transition when an in-plane electric field is applied, E$_{\rm in}$.
In the case of out-plane applied electric field, E$_{\rm out}$, the two main peak are formed by transitions from the $\sigma$ to $\pi^*$ at $11.22$~eV and the $\pi$ to $\sigma^*$ at $14.26$~eV. 
These transitions are in a good agreement with literature~\cite{NATH2015691,MOHAN20121670, doi:10.1002/andp.201900306}.
The absorption coefficient is related to the imaginary part of dielectric function. The absorption coefficient of pristine BLG shows two obvious peaks 
at $4.41$ and $14.35$~eV for E$_{\rm in}$, and two peaks at $11.78$ and $14.67$~eV for E$_{\rm out}$.  In addition, it is noticed that two peaks in reflectivity of pristine BLG are found at $4.32$ and $15.55$~eV for E$_{\rm in}$ with almost the same intensity, and two peaks at $11.22$, and $15.31$~eV for E$_{\rm out}$ with different intensities.
It should be mentioned that the peak positions of pristine BLG are strongly dependent on the interlayr distance \cite{NATH2015691, PhysicaE.64.254}. The peak intensity of $\varepsilon_2$, $\alpha$, and $R$ here in both E$_{\rm in}$ and E$_{\rm out}$ directions are quite similar to previous studies when the interlayer distance is close to $4.64$~$\angstrom$, as it is considered in our work.
It is also seen that the optical response of pristine BLG becomes insignificant beyond the
energy or optical frequency of $ \approx 25$~eV irrespective of polarization state \cite{Abdullah_2014}.

The optical response of Si-BLG is significantly modulated because the band structure and the DOS are changed. The frequency dependent variation of $\varepsilon_2$ for SiC$_{15}$ and Si$_2$C$_{14}$-II is enhanced in both directions of applied electric fields especially at the low energy range. 
This is attributed to the opeining up bandgap of these two structures. A new peak in $\varepsilon_2$ at $0.9$~eV for Si$_2$C$_{14}$-II is observed in the E$_{\rm in}$ direction referring to the existing a relativity larger bandgap. Another peak in $\varepsilon_2$ at $1.75$~eV for SiC$_{15}$ in the E$_{\rm out}$ direction is seen, which is due to decreasing the energy spacing between $\sigma$ and $\pi$ bands at the $\Gamma$-point allowing transition between these two bands at lower energy. 
In addition, a red shift in the peaks positions of these two structures occur in both directions of applied electric field arising again from the opening bandgap and significant changes in the DOS.  
Consequently, the red-shift in absorption spectra and Reflectivity have also been noticed.   

In contract, the optical response of Si$_2$C$_{14}$-I is reduced. For instant, the imaginary part of dielectric function absorption coefficient are significantly reduced. This reduction may refer to the overlapping of valence band maxima and conduction band minima at $K$- and $\Gamma$-points.  
As a result, the reflectivity is enhanced in both directions of applied electric field. 
It should be noticed that there peaks positions are almost unchanged here. The controlling of optical response by dopant atoms may be interested in opto-electronic devices.

\section{Conclusion}~\label{Sec:Conclusion}

To conclude, the first principles DFT technique is employed to compute 
the structural, electronic, mechanical, thermal, and optical properties of Si-BLG.
It is observed that tuning the concentration ratio of Si dopant atoms with different configurations, the physical properties of the structures can be modulated. 
At low ratio of Si atoms doped in BLG, a modification in s and p orbitals hybridization is seen leading to a strong fracture strain at low strain ratio. These structures show a weak behavior of mechanical response. Furthermore, a small bandgap induced at low Si dopant atoms arises an intermediate Seebeck coefficient, figure of merit, and absorption coefficient. 
Increasing the ratio of Si dopant atoms when the Si atoms are doped in both layers of the BLG, a stronger mechanical response of Si-BLG is obatined in which the fracture strain is noticed at high strain. In addition, a larger bandgap due to higher ratio of Si dopant enhances thermal and optical responses of the system. If the Si dopant atoms is doped in only one layer, the valence and conduction bands are overlapped revealing low thermal and optical behaviors. 
Our present theoretical work might help to understand some thermal and optical properties of nano
structured materials and to design nano optoelectronic devices involving graphene nano structures.

\section{Acknowledgment}
This work was financially supported by the University of Sulaimani and 
the Research center of Komar University of Science and Technology. 
The computations were performed on resources provided by the Division of Computational 
Nanoscience at the University of Sulaimani.  


%
%


\scriptsize

\end{document}